\documentclass[a4paper,10pt,conference]{IEEEtran}
\IEEEoverridecommandlockouts
\usepackage{cite}
\usepackage{amsmath,amssymb,amsfonts}
\usepackage{algorithmic}
\usepackage{graphicx}
\usepackage{textcomp}
\usepackage{xcolor}
\usepackage{bm}
\usepackage{comment}
\usepackage{enumitem}
\usepackage{subfigure}
\def\BibTeX{{\rm B\kern-.05em{\sc i\kern-.025em b}\kern-.08em
    T\kern-.1667em\lower.7ex\hbox{E}\kern-.125emX}}
\begin{document}

\title{Deep Retinex Network for Estimating Illumination Colors with Self-Supervised Learning}

\author{\IEEEauthorblockN{1\textsuperscript{st} Kouki Seo}
\IEEEauthorblockA{\textit{Department of Computer Science} \\
\textit{Tokyo Metropolitan University}\\
Tokyo, Japan \\
seo-kouki@ed.tmu.ac.jp}
\and
\IEEEauthorblockN{2\textsuperscript{nd} Yuma Kinoshita}
\IEEEauthorblockA{\textit{Department of Computer Science} \\
\textit{Tokyo Metropolitan University}\\
Tokyo, Japan \\
ykinoshita@tmu.ac.jp}
\and
\IEEEauthorblockN{3\textsuperscript{rd} Hitoshi Kiya}
\IEEEauthorblockA{\textit{Department of Computer Science} \\
\textit{Tokyo Metropolitan University}\\
Tokyo, Japan \\
kiya@tmu.ac.jp}
}

\maketitle

\begin{abstract}

We propose a novel Retinex image-decomposition network
that can be trained in a self-supervised manner.
The Retinex image-decomposition aims to decompose an image
into illumination-invariant and illumination-variant components,
referred to as ``reflectance'' and ``shading,'' respectively.
Although there are three consistencies that the reflectance
and shading should satisfy,
most conventional work considers only one or two of
the consistencies.
For this reason,
the three consistencies are considered in the proposed network.
In addition, by using generated pseudo-images for training,
the proposed network can be trained with self-supervised learning.
Experimental results show that our network can decompose images
into reflectance and shading components.
Furthermore, it is shown that the proposed network can be
used for white-balance adjustment. 

\end{abstract}


\section{Introduction}

A natural image consists of the reflectance and the shading of a scene in Retinex theory \cite{retinex}.
The reflectance and the shading are an illumination-invariant component and an illumination-variant component respectively.
Retinex image decomposition aims to decompose a natural image into two such components.
To enable the decomposition, various methods have so far been proposed \cite{chien, id1_un, id2_un, id3_un, id4_un, id1, id2, id3, id4},
where most methods are based on deep neural networks (DNN).

In the Retinex decomposition, there are three premises regard to consistency:
reconstruction consistency,  reflectance consistency in terms of exposures, and reflectance consistency in terms of illumination colors.
However, most conventional methods only considers some of them.
In contrast, conventional methods \cite{id2_un, id3_un} consider all of the premises,
but their performances are limited
due to difficultly in preparing a large amount
of real data or synthetic data for training.

Several decomposition methods trained with supervised learning have been proposed \cite{id1, id2, id3, id4}.
They often use a highly-synthetic dataset or a human-labeled dataset of the real scene \cite{data1, data2, data3}.
However, such datasets are insufficient to generalize real scenes.

To solve these problems, in this paper,
we propose a novel Retinex image decomposition network that considers both the three premises and the problem with data.
For training the proposed network,
we generate pseudo images that are taken under various exposure and illumination-color conditions.
By using such training data,
the proposed network can be trained with self-supervised learning,
and difficultly in preparing a large amount of data can be overcome.
The proposed network can decompose input image $I$
into reflectance $R_I$, gray-shading $GS_I$, and single RGB vector $\bm{c}_I$ that represents an illumination color.
Shading $S_I$ including the effect of illumination color can be obtained
by multiplying outputs $GS_I$ and $\bm{c}_I$.

We evaluate the performance of the decomposition and the estimation of illumination colors in terms of mean squared error (MSE) and hue difference $\Delta H$ of CIEDE2000 \cite{ciede2000}.
Experimental results show that our network can decompose
input images, and identify illumination colors of the input images.

\section{Preliminaries}
\subsection{Retinex decomposition}
\begin{figure}[t]
  \centering
\includegraphics[keepaspectratio, scale=0.4]{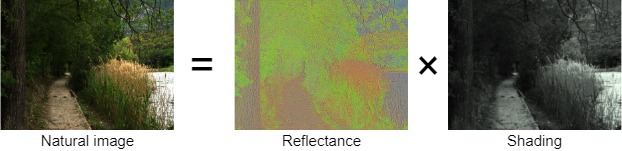}
\caption{Retinex image decomposition.\label{fig:retinex}}
\end{figure}

In Retinex theory, a natural image $I$ can be written
as the pixel-wise product of reflectance $R_I$ and shading $S_I$
as shown in Fig. \ref{fig:retinex}, i.e.,
\begin{equation}
  \label{eq:retinex}
  I(x, y) = R_I(x, y) \cdot S_I(x, y),
\end{equation}
where $(x, y)$ indicates a pixel coordinate,
$R_I(x, y)$ is in the range of $[0, 1]$,
and $S_I(x, y)$ is in the range of $[0, \infty)$.
The goal of Retinex decomposition is
to estimate reflectance $R_I$ and shading $S_I$
from a given image $I$.
Here, shading $S$ will be spatially smooth
because it is a map of the illumination intensity.
In contrast,
since $R$ is expected to include textures and edges of objects,
reflectance $R$ will be spatially discontinuous.

\subsection{Effects of exposure change on Retinex decomposition}
The change of the brightness (or exposure) of an image affects
the Retinex decomposition of the image.
Here, we discuss the effects of the exposure change.

The exposure of an image is usually expressed in terms of an exposure value (EV),
and the proper exposure for a scene is automatically decided by a camera \cite{kinoshita1, kinoshita2, kinoshita3, seo}.
The exposure value is commonly controlled by changing the shutter speed,
although it can also be controlled by adjusting various camera parameters.
Here, we assume that camera parameters except for
the shutter speed are fixed.
Let $v_0 = 0 \mathrm{[EV]}$ and $I_{v_0}$ be
the proper exposure value and the corresponding captured image
under the given conditions, respectively.
By assuming that the camera response is linear
with respect to the light intensity,
an image $I_{v_i}$ exposed at $v_i \mathrm{[EV]}$ is written as
\begin{equation}
  I_{v_i}(x, y) = 2^{v_i} I_{v_0}(x, y).
  \label{eq:exposure}
\end{equation}

From Eqs. (\ref{eq:retinex}) and (\ref{eq:exposure}),
the Retinex decomposition of $I_{\mathrm{EV}=v_0}$ and $I_{\mathrm{EV}=v_i}$
are given as
\begin{align}
  I_{v_0}(x, y) &= R_{I_{v_0}}(x, y) \cdot S_{I_{v_0}}(x, y), \\
  I_{v_i}(x, y) &= R_{I_{v_i}}(x, y) \cdot S_{I_{v_i}}(x, y) \nonumber \\
  &= 2^{v_i} R_{I_{v_0}}(x, y) \cdot S_{I_{v_0}}(x, y),
\end{align}
respectively.
Since the scenes of images $I_{v_0}$ and $I_{v_i}$
are the same,
we can obtain the following relations:
\begin{align}
  R_{I_{v_i}} &=  R_{I_{v_0}}, \label{eq:exposure_reflectance}\\
  S_{I_{v_i}} &= 2^{v_i} S_{I_{v_0}}.
\end{align}

\subsection{Effects of illumination color on Retinex decomposition}
Similarly to the exposure change,
the change of illumination color also affects shading $S_I$.

Let $\bm{c}_0 = (1, 1, 1)$, $I_{\bm{c}_0}$, and $S_{I_{\bm{c}_0}}$
be the white illumination color, 
an image taken under the illumination, and its corresponding shading,
respectively.
Then, shading $S_{I_{\bm{c}_j}}$ corresponding to $I_{\bm{c}_j}$
taken under illumination color $\bm{c}_j = (r, g, b)$ is given as
\begin{equation}
  S_{I_{\bm{c}_j}}(x, y) = \mathrm{M}_{\bm{c}_j} S_{I_{\bm{c}_0}}(x, y),
\end{equation}
where $\mathrm{M}_{\bm{c}_j} = \mathrm{diag}(\bm{c}_j)$.
For this reason,
the Retinex decomposition of $I_{\bm{c}_0}$ and $I_{\bm{c}_j}$ is given by
\begin{align}
  I_{\bm{c}_0}(x, y) 
    &= \mathrm{diag}(R_{I_{\bm{c}_0}}(x, y))
    S_{I_{\bm{c}_0}}(x, y), \\
  I_{\bm{c}_j}(x, y) 
    &= \mathrm{diag}(R_{I_{\bm{c}_j}}(x, y))
    S_{I_{\bm{c}_j}}(x, y) \nonumber \\
    &= \mathrm{diag}(R_{I_{\bm{c}_0}}(x, y))
    \mathrm{M}_{\bm{c}_j} S_{I_{\bm{c}_0}}(x, y) \nonumber \\
    &= \mathrm{M}_{\bm{c}_j} \mathrm{diag}(R_{I_{\bm{c}_0}}(x, y))
    S_{I_{\bm{c}_0}}(x, y),
\end{align}
where we used the relation
\begin{align}
  R_{I_{\bm{c}_j}} = R_{I_{\bm{c}_0}}.
  \label{eq:wb_reflectance}
\end{align}
Therefore, the relationship between $I_{\bm{c}_0}$ and $I_{\bm{c}_j}$ is written as
\begin{equation}
  \label{eq:retinex_wb}
  I_{\bm{c}_j}(x, y) 
  = \mathrm{M}_{\bm{c}_j} I_{\bm{c}_0}(x, y).
\end{equation}

\subsection{Scenario \label{subsec:scenario}}
In the Retinex decomposition, there are three premises:
\begin{description}[leftmargin=*]
  \item[Reconstruction consistency]
    The product of estimated reflectance and shading matches the corresponding original image,
    as shown in Eq. (\ref{eq:retinex}).
  \item[Reflectance consistency (exposure)]
    Reflectances are invariant against a change of exposure values,
    as in Eq. (\ref{eq:exposure_reflectance}).
  \item[Reflectance consistency (color)]
    Reflectances are invariant against a change of illumination colors,
    as in Eq. (\ref{eq:wb_reflectance}).
\end{description}

Most conventional work considers only a part of these premises,
e.g., reconstruction consistency and reflectance consistency (exposure).
In such a case,
their reflectance components are affected by the effects of exposure and
illumination-color conditions.
In literature \cite{id2_un}, all three premises are considered by training a DNN
by using videos taken by a fixed-point camera.
However, the DNN still has a limited performance
due to a limited amount of real data for training.

For these reasons, in this paper,
we propose a novel DNN for the Retinex decomposition considering all three premises and the problem with data.
For training our network,
we generate pseudo images from original images,
which correspond to images taken under various exposure and illumination-color conditions.
By using them, our network can be trained with self-supervised learning while considering above three premises.
In addition, since we generate pseudo images from general datasets,
the problem with a amount of data can be overcome.




\section{Proposed Retinex Network}
In this paper,
we aim to decompose image $I$ into reflectance $R_{I}$ and shading $S_{I}$
by using a deep neural network.
The key idea of our approach is to consider the three premises in section \ref{subsec:scenario}.
The proposed network can be trained in a self-supervised manner,
while satisfying the premises.

\subsection{Network architecture}
\begin{figure}[t]
  \centering
  \includegraphics[keepaspectratio, scale=0.35]{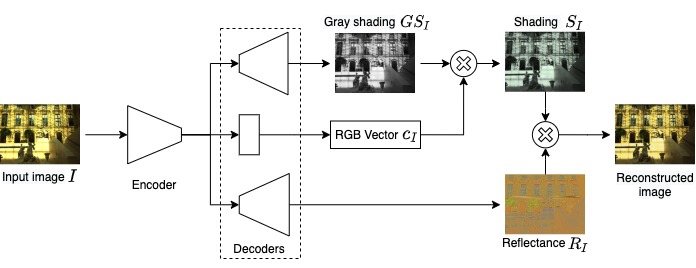}
  \caption{Network architecture \label{fig:network}}
\end{figure}

Figure \ref{fig:network} illustrates the architecture of the proposed network.
The proposed network receives input image $I$,
and outputs reflectance $R_{I}$, gray-shading $GS_I$,
and RGB vector $\bm{c}_I$.
Our network has a single encoder and three decoders.
By the encoder, input image $I$ is transformed into feature maps
that will be fed into decoders.
Reflectance $R_{I}$ with RGB color channels
is directly obtained as the output of a decoder.
In contrast, shading $S_{I}$ is given as the product of
RGB vector $\bm{c}_I$ and gray-scale shading $GS_I$.
The gray-scale shading and the RGB vector are outputted from the other two decoders, respectively.

\subsection{Data generation for self-supervised learning}
\label{subsec:datageneration}

In order to consider the three premises
in Section \ref{subsec:scenario},
images of a single scene taken under various exposure and
illumination-color conditions are required
for training the proposed network.
However, it is very costly to collect such images.
For this reason, we generate pseudo images from raw images
and use them for training the proposed network.

Because raw images are not affected by the non-linear camera response of a camera,
multiplying their pixel values by a scalar value corresponds to
the exposure change in Eq. (\ref{eq:exposure}).
In addition,
Eq. (\ref{eq:retinex_wb}) is equivalent to
applying a color-transfer matrix, used in
a white-balance adjustment in the RGB color space,
to an image.
Hence, images generated from raw images
in accordance with Eqs. (\ref{eq:scala}) and (\ref{eq:ctrans})
can be used for training the proposed network.

We utilize three color-transferred multi-exposure images
$I_{v_i, \bm{c}_i} (i \in \{1,2,3\})$
having exposure value $v_i$ and illumination color $\bm{c}_i$
for calculating loss.
Images $I_{v_i, \bm{c}_i}$ are generated
from a raw image $I_{\mathrm{raw}}$ as follows:
\begin{enumerate}
  \item Obtain an RGB image $I_{\mathrm{RGB}}$
    by demosaicing a raw image $I_{\mathrm{raw}}$.
  \item Generate three multi-exposure images $I_{v_i}$
    ($v_i \in \{-1, 0, 1\}$ [EV]) from $I_{\mathrm{RGB}}$
    in accordance with Eq.(\ref{eq:exposure}) as
    \begin{equation}
    \label{eq:scala}
      I_{v_i} =
      2^{v_i}\frac{0.18}{g(I_{\mathrm{RGB}})} I_{\mathrm{RGB}},
    \end{equation}
    where $g(I_{\mathrm{RGB}})$ indicates
    the geometric mean of the luminance of $I_{\mathrm{RGB}}$.
  \item Generate color-transferred multi-exposure images
    $I_{v_i, \bm{c}_i}$
    by multiplying $I_{v_i}$ by $M_{c_i} = \mathrm{diag}(\bm{c}_i)$
    as
    \begin{equation}
    \label{eq:ctrans}
      I_{v_i, c_i} = M_{c_i} I_{v_i},
    \end{equation}
    where $\bm{c}_i$ is a random vector in $[0.9, 1.1]^3$.
\end{enumerate}


\subsection{Loss functions}
To fulfill above premises, our network is trained to minimize the following loss function
\begin{align}
  \label{eq:all_loss}
  {\cal L} = {\cal L}_{\mathrm{recon}}
  + {\cal L}_{\mathrm{reflect}} + {\cal L}_{\mathrm{other}},
\end{align}
where ${\cal L}_{\mathrm{recon}}$ is the image-reconstruction loss between input images and reconstructed ones.
${\cal L}_{\mathrm{reflect}}$ and ${\cal L}_{\mathrm{other}}$ are loss functions for constraining outputs $\hat R_{I_i}$, and $\hat S_{I_i}$ and $\hat c_{I_i}$, respectively.

For the reconstruction consistency,
in accordance with Eq.(\ref{eq:retinex}), 
we use image-reconstruction loss ${\cal L}_{\mathrm{recon}}$
so that the pixel product of $\hat R_{I_i}(x, y)$ and $\hat S_{I_i}(x, y)$ is 
equal to input image $I_i \triangleq I_{v_i, c_i}$.
We calculate ${\cal L}_{\mathrm{recon}}$
for all combinations of the input images and the pixel product $\hat R_{I_i}(x, y) \cdot \hat S_{I_i}(x, y)$
as
\begin{align}
\label{eq:recon_loss}
{\cal L}_{\mathrm{recon}} &= \sum_{i=1}^3 \sum_{j=1}^3 
\{ \lambda_1 \| I_i(x, y) - \hat R_{I_j}(x, y) \cdot \hat S_{I_i}(x, y) \|^2 \nonumber \\
&+ \lambda_2 \| 1- \mathrm{SSIM} \mathit (I_i(x, y), \, \hat R_{I_j}(x, y) \cdot \hat S_{I_i}(x, y)) \|^2 \nonumber \\
&+ \lambda_3 \| \Delta E (I_i(x, y), \, \hat R_{I_j}(x, y) \cdot \hat S_{I_i}(x, y)) \|^2 \}  ,
\end{align}
where $\lambda_1, \lambda_2$ and $\lambda_3$ are weights of the loss terms,
$\|\cdot\|$ is L2 norm,
$\mathrm{SSIM} (\cdot)$ calculates a structural similarity (SSIM) value,
and $\Delta E (\cdot)$ calculates the CIEDE2000 color difference \cite{ciede2000}.
By using SSIM and $\Delta E (\cdot)$ as the loss terms,
images reconstructed by using outputs $\hat R_{I_i}$ and $\hat S_{I_i}$ reproduce
the details of input images $I_i$,
and moreover output reflectance $\hat R_{I_i}$ can be consistent regardless of exposure and illumination-color conditions.

Also, we use reflectance loss ${\cal L}_{\mathrm{reflect}}$ to improve the consistency of output reflectance $\hat R_{I_i}$ as
\begin{align}
\label{eq:r_loss}
{\cal L}_{\mathrm{reflect}} = \sum_{i=1}^3 \sum_{j=1}^3
\{ \lambda_4 \| \hat R_{I_i}(x, y) - \hat R_{I_j}(x, y) \|^2 \nonumber \\
+ \lambda_5 | 0.5 - \mathrm{mean} \mathit (\hat R_{I_i}) | \} ,
\end{align}
where $\lambda_4$ and $\lambda_5$ are weights of the loss terms,
$\mathrm{mean} (\cdot)$ calculates the mean value of the whole reflectance.
By adjusting the mean value to $0.5$, 
our network can output the normalized color information of input images $I_i$ as reflectance $\hat R_{I_i}$.

To add smoothness to output shading $\hat S_{I_i}$,
the total variation $\mathrm{tv}(\cdot)$ is utilized as a loss function for shading.
Combining $\mathrm{tv}(\cdot)$ and a loss function of output RGB vector $\hat c_{I_i}$,
we calculate ${\cal L}_{\mathrm{other}}$ as follows:
\begin{equation}
\label{eq:other_loss}
{\cal L}_{\mathrm{other}} = \sum_{i=1}^3
\{ \lambda_6 \, \mathrm{tv} \mathit (\hat S_{I_i}) + \lambda_7 \| c_i - \hat c_{I_i} \|^2 \},
\end{equation}
where $\lambda_6$ and $\lambda_7$ are weights of the loss terms.

In practice, we empirically set $\lambda_1 =3, \lambda_2 =1, \lambda_3 =2, \lambda_4 =3, \lambda_5 =1, \lambda_6 =10$ and $\lambda_7 =20$ as weights, respectively.

\section{Simulation}
We performed two simulations to confirm the performance of the proposed network.
For training our network, we used 3640 raw images in the HDR+ Burst Photography Dataset \cite{dataset1}.

\subsection{Result of Retinex Decomposition}
\begin{figure*}[t]
  \centering
\includegraphics[keepaspectratio, scale=0.3]{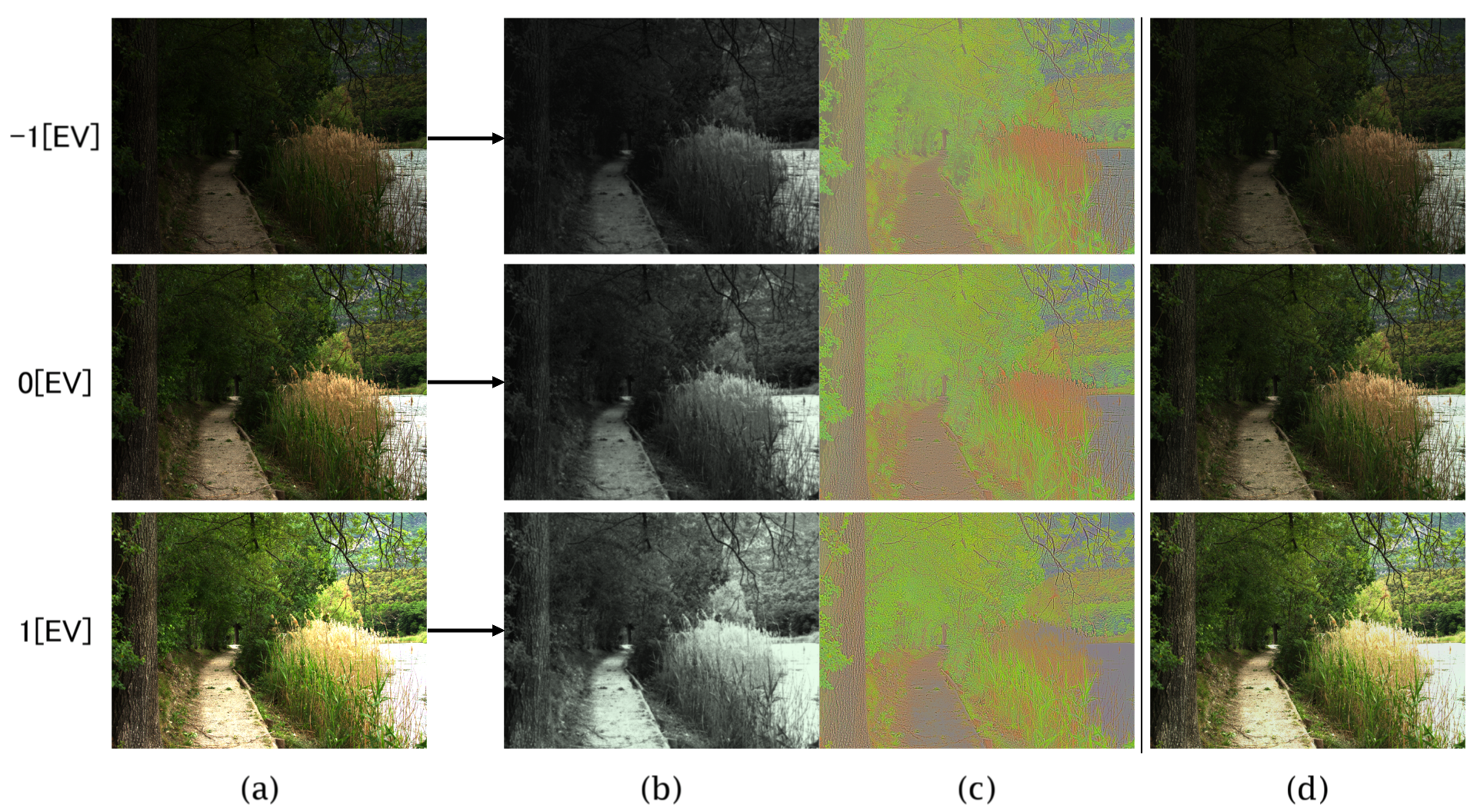}
\caption{Example of images generated by our network.
(a) Input image. (b) Output shading. (c) Output reflectance. (d) Reconstructed image by using output components.
\label{fig:result}}
\end{figure*}

Figure \ref{fig:result} shows
an example of images outputted from our network as Retinex decomposition and reconstruction.
From Fig.\ref{fig:result},
our network was confirmed to generate almost the same reflectance from three input images with different exposures.
Figure \ref{fig:result} also shows
that the input images with different exposures were able to be reconstructed by using output components.
From these results, our network was demonstrated to work well.

\subsection{Result of white-balance adjustment}
\begin{figure}[t]
  \centering
  \includegraphics[keepaspectratio, scale=0.355]{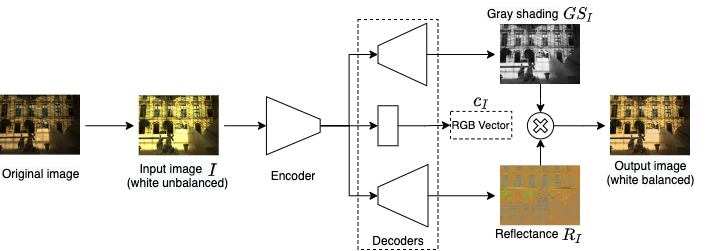}
  \caption{WB adjustment with our network \label{fig:simulation}}
\end{figure}

To evaluate the estimation performance of illumination colors,
a WB adjustment was applied to input images,
where the input images were prepared as white unbalanced images by using only color-transferring,
i.e. using steps (1) and (3) in Sec.\ref{subsec:datageneration}.
Figure \ref{fig:simulation} shows the process of the WB adjustment used in this experiment.
In the process,
outputted RGB vectors were not used for reconstructing output images
so that the effects of the illumination color included in input images were eliminated from the images.



In this experiment,
100 color-transferred images,
which were generated from 100 raw images in the RAISE Dataset \cite{dataset2},
were applied to the trained network as input images. 
Output images produced from the proposed network were evaluated
in terms of MSE and hue difference $\Delta H$ of CIEDE2000 \cite{ciede2000}.
To confirm the decomposition performance of our network,
the scores of output images were compared with those of input ones,
where original images that were not color-transferred were used as reference ones for calculating scores. 

\begin{table}[t]
  \centering
\caption{Scores of WB adjustment simulation.\label{tab:result}}
\begin{tabular}{cccll}
\cline{1-3}
                     & MSE                  & $\Delta H$       &  &  \\ \cline{1-3}
Input                & 0.0259               & 3.5017           &  &  \\
Output               & {\bf 0.0198}          & {\bf 3.2403}    &  &  \\ \cline{1-3}
\multicolumn{1}{l}{} & \multicolumn{1}{l}{} & \multicolumn{1}{l}{} &  & 
\end{tabular}
\end{table}
\begin{figure}[t]
    \centering
	\subfigure[Original image (Reference)]{
		\includegraphics[width = 0.4\columnwidth]{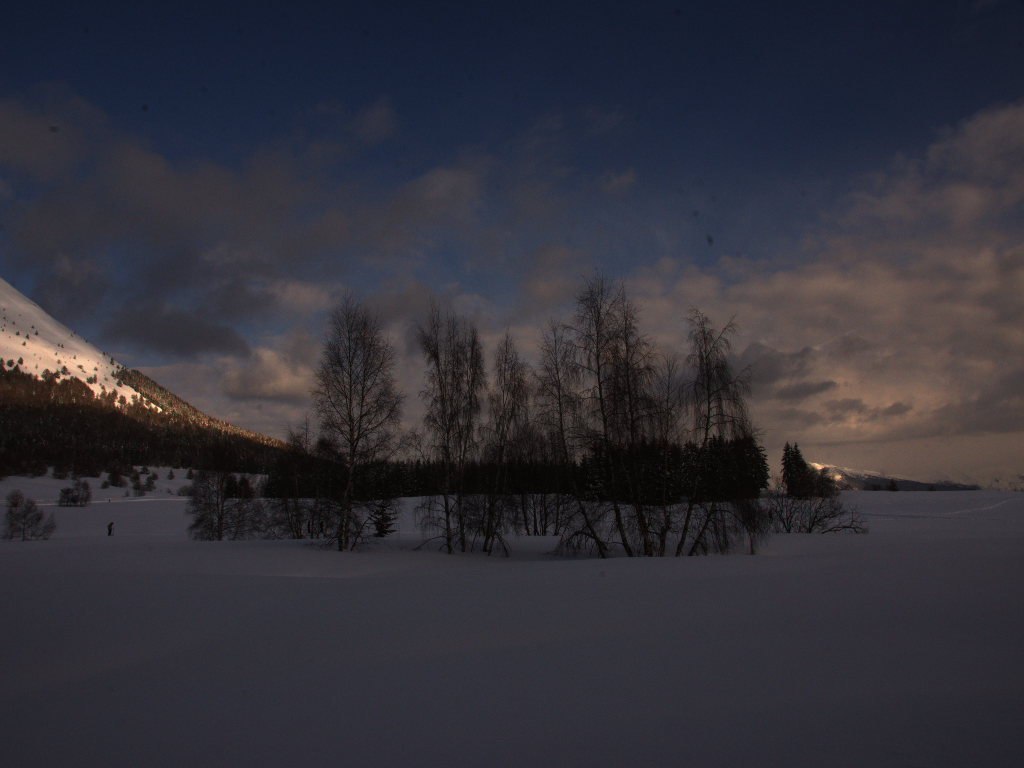}
		\label{subfig:ref}} \\
    \subfigure[Input image]{
		\includegraphics[width = 0.4\columnwidth]{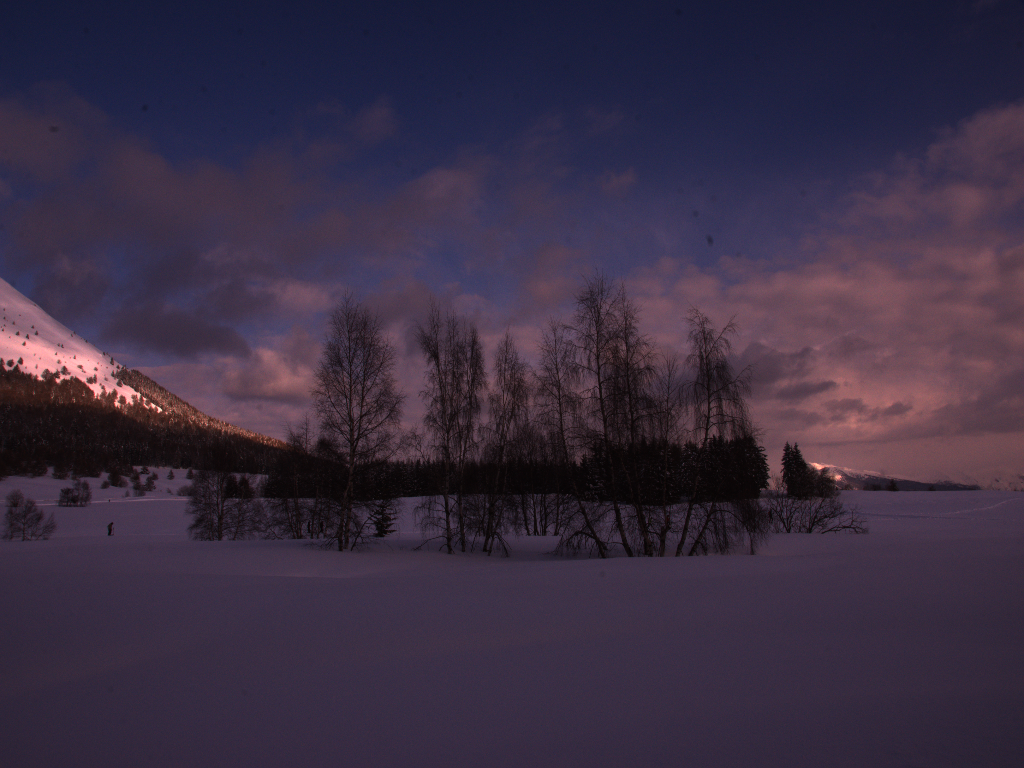}
		\label{subfig:input}}
	\subfigure[Output image]{
		\includegraphics[width = 0.4\columnwidth]{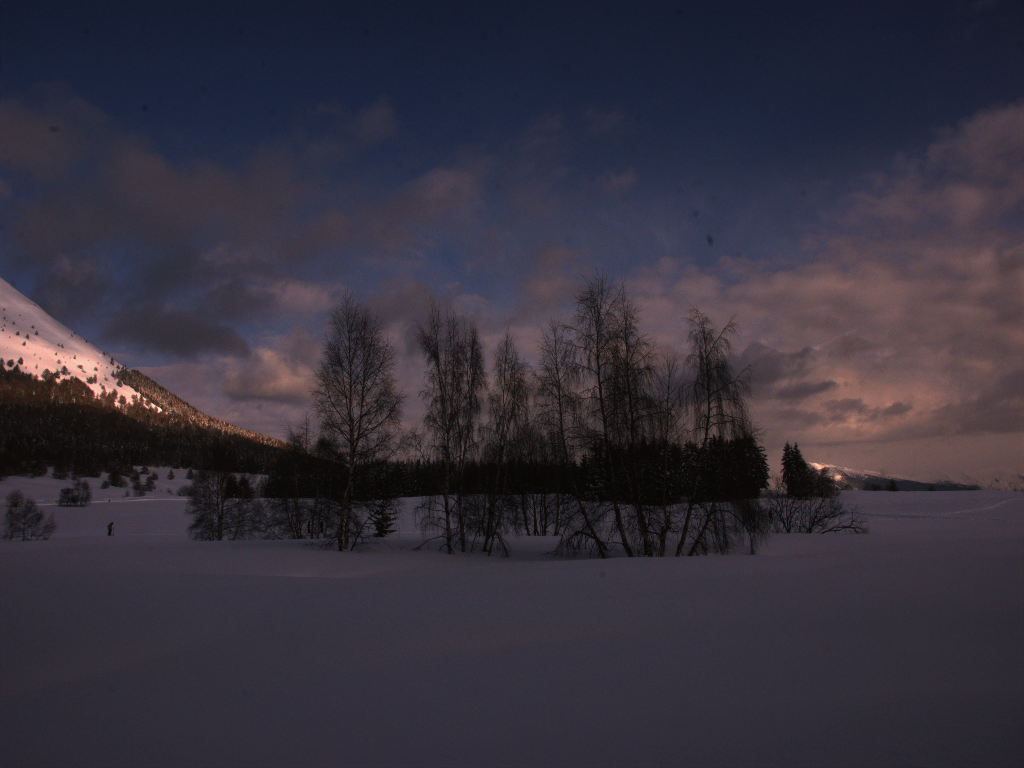}
		\label{subfig:output}}
    \caption{Example of WB adjustment with out network \label{fig:result_img2}}
\end{figure}
Table \ref{tab:result} shows the scores of MSE and hue difference $\Delta H$,
which were averaged over all 100 images.
From Table \ref{tab:result},
both scores of the output images were lower than those of the input images,
where a smaller value indicates a better result in the scores.
Figure \ref{fig:result_img2} shows an example of the reference, input, and output images used in this experiment.
From Fig.\ref{fig:result_img2},
the white balance of the output image was closer to the reference one than the input one.
Therefore, our network was confirmed to be able to eliminate the effects of the illumination color of the input image.

\section{Conclusion}
In this paper, we proposed a novel Retinex image decomposition network
considering the premises of the Retinex decomposition.
In addition, the proposed network can be trained in a self-supervised manner
by using pseudo-generated images with various exposures and illumination colors.
In an experiment,
our network was demonstrated to be able to generate almost the same reflectance from input images with different exposures
and estimate illumination colors.

\end{document}